\documentclass[aps,prb,reprint,superscriptaddress,twocolumn]{revtex4}

\usepackage{amsfonts}
\usepackage{amsmath}
\usepackage[normalem]{ulem}
\usepackage{bm}
\usepackage{graphicx}
\usepackage{subfig}
\usepackage{lipsum}
\usepackage{array,multirow}
\usepackage[utf8x]{inputenc}
\usepackage{hyperref}

\def\opone{\leavevmode\hbox{\small1\kern-3.8pt\normalsize1}}
\usepackage{color}

\begin{document}
	
	\title{Measurement of the Thulium Ion Spin Hamiltonian Within a Yttrium Gallium Garnet Host Crystal}
	
	\author{Jacob H. Davidson}
	\affiliation{Qutech and Kavli Institute of Nanoscience, Delft University of Technology, Delft, The Netherlands}
	%-----------------------------
	\author{Philip J.T. Woodburn}
	\affiliation{Department of Physics, Montana State University, Bozeman, Montana 59717, USA}
	%-----------------------------
	\author{Aaron D. Marsh}
	\affiliation{Department of Physics, Montana State University, Bozeman, Montana 59717, USA}
	%-----------------------------
	\author{Kyle J. Olson}
	\affiliation{Department of Physics, Montana State University, Bozeman, Montana 59717, USA}
	%-----------------------------
	\author{Adam Olivera}
	\affiliation{Department of Physics, Montana State University, Bozeman, Montana 59717, USA}
	%-----------------------------
	\author{Antariksha Das}
	\affiliation{Qutech and Kavli Institute of Nanoscience, Delft University of Technology, Delft, The Netherlands}
	%-----------------------------
	\author{Mohsen Falamarzi Askarani}
	\affiliation{Qutech and Kavli Institute of Nanoscience, Delft University of Technology, Delft, The Netherlands}
	%-----------------------------
	\author{Wolfgang Tittel}
	\affiliation{Qutech and Kavli Institute of Nanoscience, Delft University of Technology, Delft, The Netherlands} 
	%-----------------------------
	\author{Rufus L. Cone}
	\affiliation{Department of Physics, Montana State University, Bozeman, Montana 59717, USA}
	%-----------------------------
	\author{Charles W. Thiel}
	\affiliation{Department of Physics, Montana State University, Bozeman, Montana 59717, USA}
	
	%\date{\today}
	
	%-----------------------------	

	\begin{abstract}
 We characterize the magnetic properties for thulium ion energy levels in the Y$_3$Ga$_5$O$_{12}$ (Tm:YGG) lattice with the goal to improve decoherence and reduce line-width broadening caused by local host spins and crystal imperfections. More precisely, we measure hyperfine tensors for the lowest level of the, $^3$H$_6$, and excited, $^3$H$_4$, states using a combination of spectral hole burning, absorption spectroscopy, and optically detected nuclear magnetic resonance. By rotating the sample through a series of angles with an applied external magnetic field, we measure and analyze the orientation dependence of the Tm$^{3+}$ ion's spin-Hamiltonian. Using this spin-Hamiltonian, we propose a set of orientations to improve material properties that are important for light-matter interaction and quantum information applications. Our results  yield several important external field directions: some to extend optical coherence times, another to improve spin inhomogeneous broadening, and yet another that maximizes mixing of the spin states for specific sets of ions, which allows improving optical pumping and creation of lambda systems in this material. 

	\end{abstract}
	
\maketitle

\section{Introduction}

 Rare-earth-ion-doped crystals (REIC) have long been studied for use in high frequency quantum and classical signal processing \cite{Babbitt2014, Tittel2010}. The transitions of interest are those between states of the 4f$^N$ configuration that are weakly permitted due to weak mixing with opposite parity states by the host crystal field. Additionally, these electrons are spatially isolated behind the 5s and 5p shell electrons that are distributed farther from the nucleus. These transitions have shown optical and spin coherence times of up to hundreds of microseconds and minutes, respectively, that enable interesting applications in quantum and classical data processing \cite{Sun2005,Kaplyanskii2012}. This coherence, and the associated applications, are often limited by time-varying  electromagnetic fields in the local crystalline environment \cite{Fraval2004,Macfarlane1980}.

Trivalent thulium has received particular interest due to its simple energy level structure, long optical coherence time, and the availability of diode lasers at the 795 nm optical transition wavelength\cite{Guillot-Noel2005,Ohlsson2003}. Of particular interest is Tm$^{3+}$ embedded in Yttrium Gallium Garnet (Y$_3$Ga$_5$O$_{12}$, Tm:YGG) for which optical coherence times ($T_2$) on the order of milliseconds have recently been measured \cite{Thiel2014PRB,Thiel2014PRL}. This ion and host crystal combination is still relatively unexplored compared to Tm:YAG. %For example, the difference in coherence times between these two structurally similar, yet chemically different, garnet materials is not yet completely understood \cite{Sun2005, Thiel2014PRB}.
However, deeper understanding about the crystal dynamics, symmetry, and hyperfine structure has led to huge improvements in spectroscopic properties of optical and microwave transitions in other rare-earth-doped materials. One such example is the possibility of transitions with zero first order Zeeman shifts (ZEFOZ), which may be obtained by controlling the interplay of the measured terms of the ion's effective spin Hamiltonian, that can lead to orders of magnitude increase in spin coherence times \cite{Fraval2004}. Additionally, an effective spin Hamiltonian approximation can be used to improve and develop new experiments for applications of quantum networking and light matter interaction \cite{Zhong2015,Rancic2018}.

In this letter, we present a complete characterization of the hyperfine structure in the ground and excited states of the 795nm optical transition of Tm$^{3+}$ ions. To perform these measurements, we use absorption spectroscopy, spectral hole burning (SHB), and optically detected nuclear magnetic resonance (ODNMR) to isolate, measure, and confirm different terms of the effective spin Hamiltonian. These measurements individually produce information about the coupled behavior of these levels and the six magnetically equivalent, orientationally in-equivalent sites. Only by cross referencing the spectra from different sets of measurements are we able to decouple the information into the individual orientational behavior of each thulium ion in each individual site for each electronic level. Our results are then confirmed to be in good agreement with initial assumptions about crystalline structure\cite{Thiel2014PRB} and a pair of independent measurements of the spin Hamiltonian terms using a single crystalline direction of Tm:YGG. This new knowledge allows us to predict the presence of optical clock transitions in this material that may allow extending the optical coherence times for light storage and  quantum memory.

This letter is presented in the following order: In Sec.\ref{section2}, we give a brief overview of the Hamiltonian of the system and the remaining unknowns to be measured. In Sec.\ref{section3}, we describe the general crystal geometry as well as the implications with regard to the Hamiltonian and how to isolate the specific projections of an external field on a specific ion site. Section \ref{section4} describes the experimental setups used to produce the results. In Secs. \ref{section5} and \ref{section6}, we describe how we gather spectra from hole burning and ODNMR measurements with different crystal orientations and which hyperfine tensor values stem from the results.  Section \ref{section7} contains details on a pair of independent measurements that confirm these tensor values for a specific orientation of the external field. In Sec. \ref{section8}, we discuss our results in the context of optical clock transitions and we identify a series of field orientations that may lead to improved optical coherence in this material.

\section{Spin Hamiltonian for the Enhanced Zeeman and Quadratic Zeeman effects} \label{section2}

For a non-Kramers ion contained in a crystal lattice of low symmetry, a six term Hamiltonian is often considered to describe the behavior of the ionic level structure \cite{Macfarlane1987,Abragam2012}.
\begin{equation}
H= (H_{FI} + H_{CF}) + H_{HF} + H_{Q}  + H_{eZ} + H_{nZ} 
\end{equation}
The first pair of terms describes the energy structure of a free ion ($H_{FI}$) coupling to the local crystal field ($H_{CF}$), which determines the energies of the states differentiated by total electron spin angular momentum that we are interested in for this work. The remaining four terms from contributions of the hyperfine ($H_{HF}$), quadrupole ($H_{Q}$), electronic Zeeman($H_{eZ}$) and nuclear Zeeman ($H_{nZ}$) interactions are treated as small perturbations. The measurements conducted here focus on a specific pair of these electronic levels, specifically the lowest energy crystal field levels of the  $^3$H$_6$ and $^3$H$_4$ multiplets.  Additionally, the only isotope of thulium ($^{169}$Tm) possesses a nuclear spin of $\frac{1}{2}$ so that there is no nuclear quadrupole splitting and the $H_{Q}$ term can be ignored \cite{Teplov1967,Guillot-Noel2005,Veissier2016}.

The remaining terms of this Hamiltonian are magnetic in nature, corresponding to the hyperfine, electronic Zeeman, and nuclear Zeeman interactions. By expanding these remaining terms to second order as in Ref[15]\cite{Teplov1967} we arrive at a three term Hamiltonian,
\begin{equation}
H=\boldsymbol{B}\cdot (-g_n \beta_n \boldsymbol{\mathbb{I}} - 2g_J \mu_B A_J \boldsymbol{\Lambda}_J)\cdot \boldsymbol{I} - g_J^2 \mu_B^2 \boldsymbol{B} \cdot \boldsymbol{\Lambda}_J \cdot \boldsymbol{B}
 \label{Ham}
\end{equation}
where $g_n$ is the nuclear gyro-magnetic ratio of thulium, $\beta_n$ is the nuclear magneton, $g_J $ is electronic g factor for each level,  $\mu_B$ is the Bohr magneton, $A_J$ is the hyperfine interaction constant, $\boldsymbol{I}$ the ion's nuclear spin, $\boldsymbol{B}$ the applied external field, and, finally, $\boldsymbol{\Lambda}_J$ the hyperfine tensor. This tensor is given by second order perturbation theory as
\begin{equation}
\Lambda_{J,(\alpha, \beta)}=\sum_{n \neq 0} \frac{\langle 0 |J_{\alpha}| n\rangle \langle n | J_{\beta} | 0\rangle}{E_n-E_0}, \quad \quad \alpha,\beta \in\{x,y,z\}
\label{Lambda}
\end{equation} 
where  $\langle 0 |J_{\alpha}| n\rangle, \langle n | J_{\beta} | 0\rangle$ are wave-function matrix elements for the electronic levels of interest, and the sum is over the crystal field levels with energies $E_n$ \cite{Teplov1967,Guillot-Noel2005,Veissier2016}.

The first two terms in Eq.\ref{Ham} describe the enhanced effective nuclear Zeeman effect that results from the second order coupling of the hyperfine interaction between the electronic Zeeman effect and the first order nuclear Zeeman interaction. The third term is the quadratic Zeeman effect that results from the expansion of the electronic Zeeman term where the orientation dependence is again determined by a scaled version of the same hyperfine tensor, $\boldsymbol{\Lambda}$ \cite{Bleaney1982}. Other terms that arise from the expansion to second order are small and neglected.  With this description in place, the only unknowns in Eq.\ref{Ham} are the 18 matrix elements of the hyper-fine tensors $\boldsymbol{\Lambda_{g,e}}$, with each a 3x3 matrix that describes the magnetic interactions of these levels for Tm$^{3+}$ ions in this material. The level structure with an applied magnetic field is shown in Fig. \ref{SHBFig} a. The effect of the quadratic term can be seen as a shift of the whole electronic crystal field level, and the linear term (\ref{SHBFig} a.\textbf{Inset}) splits the pair of nuclear spin states. The number of unknown elements in this tensor drops further due to the local symmetry of the Tm$^{3+}$ ions that occupy yttrium in the YGG lattice.

\section{Crystal Symmetry and Site Selection} \label{section3}

\begin{figure}[t]
\begin{centering}
\includegraphics[width=0.5\textwidth]{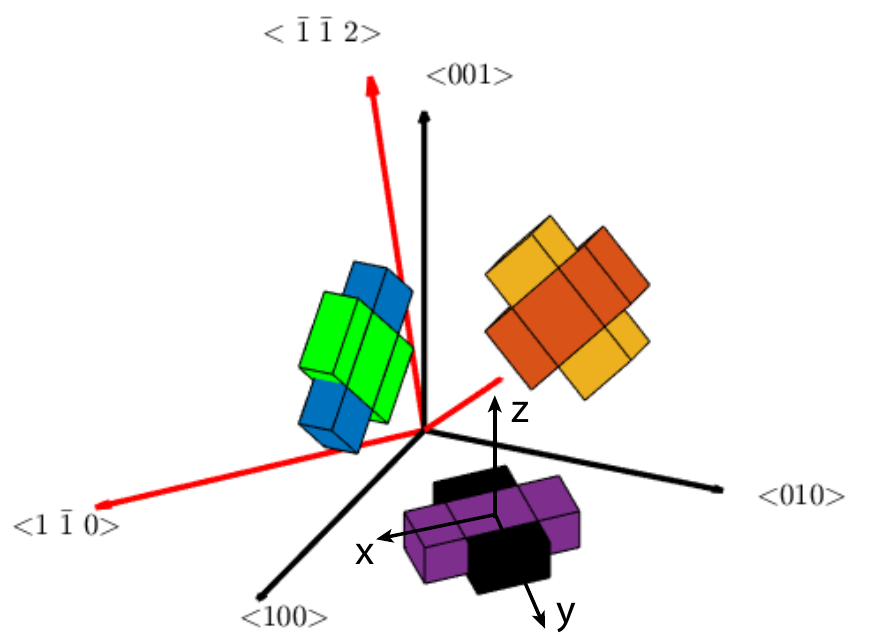}	
\caption {Possible sites of Tm$^{3+}$ ions in a YGG crystal. Black thick axes denote the cubic unit cell boundaries. Each of the 6 colored cuboids denotes the location as well as the direction of the ion's local Y-axis (along the cuboids largest dimension),that is also expected to be the direction of the optical transition dipole moment of the thulium transition, relative to a cubic crystal cell. Each site is identified by a specific color throughout the remainder of this letter: "Site 1" (Black), "Site 2" (Purple), "Site 3" (Orange), "Site 4" (Yellow), "Site 5" (Green), "Site 6" (Blue). Red axes denote the symmetry axes along which the crystal is grown and oriented for these studies. Throughout this letter we will characterize rotations with respect to these directions $<1\bar{1}0>,<111>,<\bar{1}\bar{1}2>$.  Finally, the labeled (x,y,z) thin black axes denote the local site axes as defined by the D$_2$ point group symmetry(shown for the example of Site 1).}
\label{CrystalCell}
\end{centering}
\end{figure}

The location and symmetry of the yttrium sites is well understood for garnet crystals\cite{Dillon1961, Menzer1929, Sun2005}. YGG is a cubic crystal with space group symmetry O$^{10}_{h}$, and the thulium sites feature D$_2$ point group symmetry\cite{Thiel2014PRB}. The site symmetry, as described for the similar case of Tm:YAG\cite{Sun2000} and depicted in Fig. \ref{CrystalCell}, allows for six crystallographically equivalent yet orientationally in-equivalent substitution sites.

The magnetic behavior  of Tm$^{3+}$ in each of these sites is the same, due to the ion being locally surrounded by the same structural and chemical environment. However, the ion and its local environment occurs with six different orientations with respect to the cubic crystal cells of the garnet, leading to orientational, and in essence magnetic, in-equivalence between the sites.  Here, as in Ref.[20]\cite{Sun2000} and Fig \ref{CrystalCell}, there are specific directions with respect to the cubic crystal axes that cast different subsets of the orientationally in-equivalent sites into classes of ions that share the same projections of applied electromagnetic fields ($\boldsymbol{\vec{E}},\boldsymbol{\vec{B}}$) onto the local Tm$^{3+}$ site axes. For fields ($\boldsymbol{\vec{E}},\boldsymbol{\vec{B}}$) directed along the crystalline $<$111$>$ and $<$001$>$ sets of directions ($<$110$>$,$<\bar{1}\bar{1}2>$  and $<$100$>$,$<$010$>$ respectively) . These symmetric directions serve as a labeling convention for growth, dicing, and orientation of these crystals  and are built into the models detailed below, which allows us to determine the relative projection of incident fields on each of the Tm$^{3+}$ ion sites throughout our experiments. 

 On a more microscopic level, the local D$_2$  symmetry for each site is such that, by using a set of axes around the Tm$^{3+}$ ion defined by that site symmetry, we can diagonalize the hyperfine tensor from Sec. \ref{section2}. This leaves only the six diagonal elements, $\Lambda_{J,(x,y,z)}$(three for each electronic state  $J$), which correspond to the tensor elements along the three axes shown in Fig.\ref{CrystalCell} for "Site 1" of the six sites. Considering the crystal symmetry, the entire enhanced effective Zeeman term of Eqn. \ref{Ham} for a given electronic level can be simplified to \cite{Guillot-Noel2005,Veissier2016} 
\begin{equation}
H_{J,(x,y,z)}= -\hbar\sqrt {g_{J,x}^2B_x^2+ g_{J,y}^2B_y^2+ g_{J,z}^2B_z^2} \cdot \boldsymbol{I} 
\label{splittingEqn}
\end{equation}
for
\begin{equation}
g_{J,(x,y,z)}=  -g_n \beta_n - 2g_J \mu_B A_J \Lambda_{J,(x,y,z)}.
\label{gvalEqn}
\end{equation}
The determination of the six remaining parameters (three per electronic level) will be discussed in later sections.

 To measure these parameters, we gather data on the splittings of the ground and excited hyperfine states for a series of different orientations of a Tm:YGG crystal sample. In spherical coordinates, each specific orientation ($\theta,\phi$) relative to the cubic crystal cell shown in Fig.\ref{CrystalCell} has a fixed set of projections on each of the 6 potential Tm$^{3+}$ ion site axes.  Thus, for every angle of the magnetic field with a macroscopic, oriented sample, we can determine from the expected geometry the exact magnetic projection on each Tm$^{3+}$ ion axis. By matching sets of these magnetic field projections to various measurements of the hyperfine splitting we determine the desired hyperfine tensor values.

\section{Experimental Setup}\label{section4}

To gather the data for our study, we used a total of five Tm:YGG crystals, each one cut along the symmetric dimensions of the cubic crystal cell. In addition, we employed several experimental setups. While similar in concept (Laser $\rightarrow$ Pulsing $\rightarrow$ Frequency Control $\rightarrow$ Cold Crystal in a magnetic field $\rightarrow$ Detector), they featured technological differences. For example, we used different cryogenic systems --two helium flow cryostats reaching a base temperature of <2K, equipped with magnets that allow applying fields of up to 7 T and fitted with a gearing system that allows rotating the cooled sample by up to 360 degrees around a fixed optical axis for each crystal sample; and an adiabatic demagnetization refrigerator reaching 500 mK temperature, equipped with a 2T magnet as well as a static crystal holder. 

To address the center of the inhomogeneously broadened  $^3$H$_6$ $\Leftrightarrow$ $^3$H$_4$ optical transition of thulium at 795.325 nm wavelength, we used either a titanium sapphire laser or an external cavity diode laser. The light was subsequently directed through an acousto-optic modulator (AOM), which allowed creating pulses out of the continuous-wave laser light. Additionally, we used either an AOM in double-pass configuration, or a phase modulator driven by a serrodyne signal, to frequency modulate the light. The appropriately prepared pulses were then directed to the cooled Tm:YGG samples either by free space or by using a single-mode optical fiber. The beam diameter inside the samples were around 500 $\mu$m in case of the flow cryostat and free-space coupling, and around 200 $\mu$m in the other case. Finally, the light was sent into a photodetector, and the electrical signals were displayed on an oscilloscope and recorded for analysis.

The set of ODNMR experiments detailed in section \ref{section6} also required the application of radio-frequency signals that vary between 20kHz and 10 MHz. Towards this end, one of the flow cryostats is equipped with an impedance matched copper coil that provides an adjustable frequency RF field around the crystal sample. This coil is driven using 5-10 W RF power, created by means of a RF signal generator, a pair of 40dB amplifiers, as well as variable attenuation. For each frequency, the output from the signal generator was set to the desired value and mixed with RF noise, low passed to 100kHz before being sent to the coil, and terminated to 50$\Omega$. The added noise increases the RF tone bandwidth to provide an RF field capable of driving a larger population of spins, and in turn create signals with higher SNR. 

Additional information about the experimental setups will be given in Sections \ref{section5}, \ref{section6}, \& \ref{section7} along with the description of the measurements.

\section{Spectral Hole Burning Description and Results} \label{section5}

\begin{figure}[t]
\begin{centering}
\includegraphics[width=0.5\textwidth]{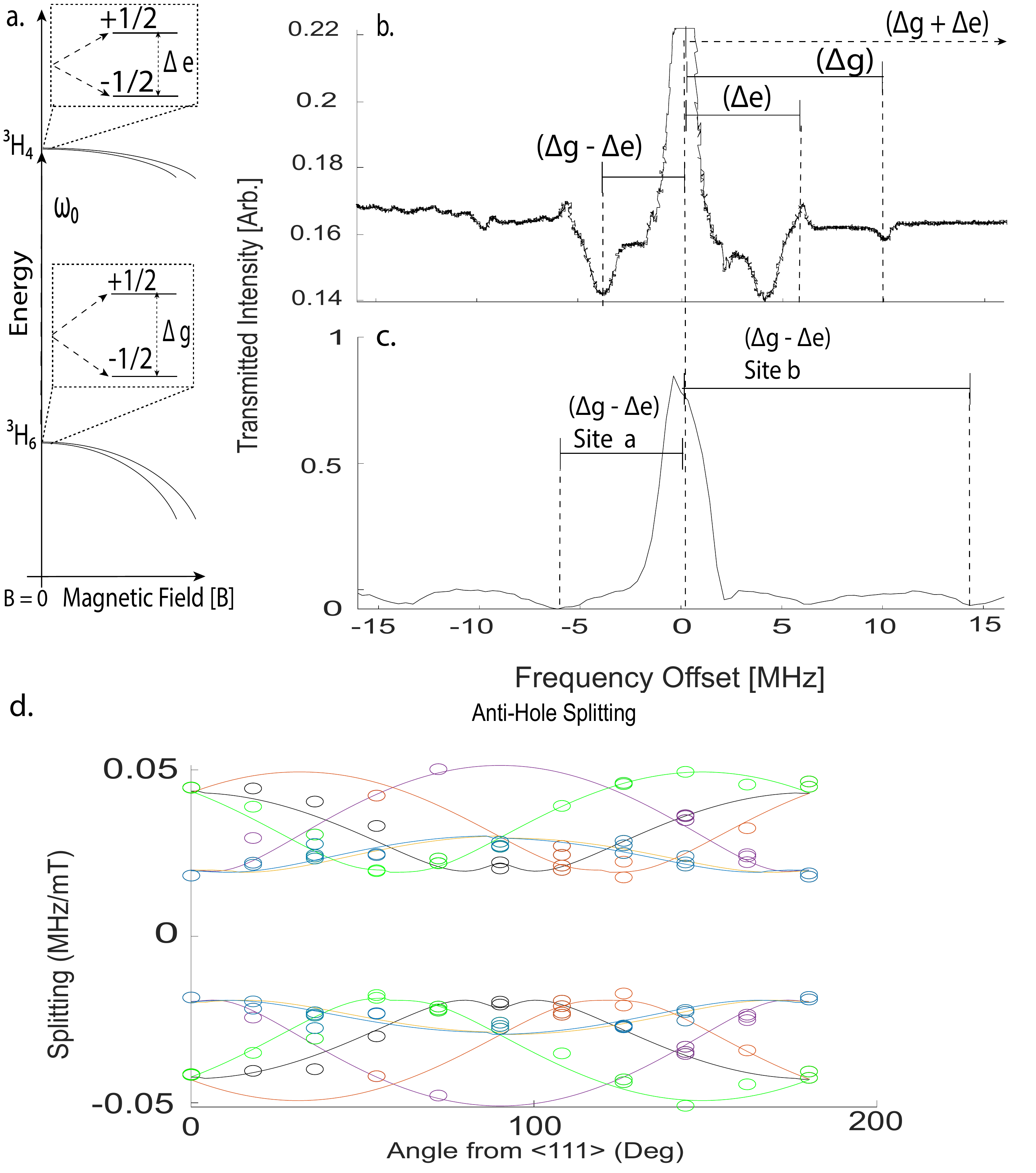}	
\caption {\textbf{a.} Diagram of the level structure for the optical transition in $^{169}$Tm$^{3+}$ with increasing magnetic field. Note that on a large scale, the change in energy of all the states is dominated by the quadratic term, but for small fields it is linear. \textbf{b.} A detailed serrodyne scan of a spectral hole structure measured for a magnetic field of 90 mT applied along the crystalline $<$111$>$ axis. The small spectral feature at $\sim$2 MHz likely arises from the only other set of sites for this orientation. \textbf{c}. Double passed AOM frequency scan of a broader spectral hole at 300mT. Only $\Delta_g- \Delta_e$ anti-holes are visible for a pair of differently oriented sites. \textbf{d}. Markers show the anti-hole frequencies plotted vs the angle between the external magnetic field and the crystal's $<$111$>$ axis. Solid lines of each color are fits of a ``difference" tensor to each of the six potential sites for the ions with a correlation coefficient R$^2$ $> 90 \%$. }
\label{SHBFig}
\end{centering}
\end{figure}

The six unknown tensor values appear in both the linear and quadratic terms of the Hamiltonian such that measurements of either term's orientation dependence is sufficient to determine the entire spin-Hamiltonian. The linear splitting of hyperfine levels the ground and excited states can be observed through spectral hole burning measurements \cite{Macfarlane1987}. With a magnetic field applied, both the ground and excited electronic states split into spin up $+\frac{1}{2}$, ($|\uparrow\rangle$) and spin down $-\frac{1}{2}$, ($|\downarrow\rangle$) states, creating a pair of levels seen in Fig.\ref{SHBFig} a. (Inset). Even at cryogenic temperatures, both ground-state spin states are equally populated. Spectral hole burning drives atomic population through this level structure, resulting in long lasting population differences between the different spin states\cite{Ohlsson2003}. The details of this optical pumping mechanism are well understood and are detailed in Ref.[21,22]\cite{Lauro2009, Louchet2007}. The initial pump pulse creates the main spectral hole: a reduction of the atomic absorption of the ensemble at the chosen pulse frequency within the in-homogeneous broadening of the crystal. 

Considering only a single projection of the magnetic field on a site, ions in the selected bandwidth are pumped to the other ground-state hyperfine level and thereby create a feature with increased absorption, referred to as an anti-hole, at a frequency offset given by the ground-state splitting, $\Delta_g$. Due to the inhomogeneously broadened nature of the transition, additional anti-holes and a single side hole arise symmetrically around the main hole at frequency offsets of ($\Delta_g$ $\pm$ $\Delta_e$) and $\Delta_e$, respectively.  By frequency sweeping weak laser light over the spectral region surrounding the main hole and recording the transmitted intensity, these population differences are visible as shown in Fig.\ref{SHBFig} b. By measuring the frequency offset of side and anti-holes for a set of different orientations we can determine the hyperfine tensor in Eq.\ref{Ham}. 

For an arbitrary orientation, each site has a different projection of the external field on its local axes. Thus, based on that projection, it is possible that that the same spectral feature(e.g. the first anti-hole at $\Delta$g - $\Delta$e) from separate sites may be visible at different frequencies. In the worst case, this could result in crowded spectra that contain as many as 48 features that would need to be assigned to different sites and associated external field projections. Fortunately, due to the fixed orientations of the sites with respect to the symmetric crystal axes, some orientations cast sites into equivalent subsets, making their spectral features appear at the same frequencies. This is the case in Fig.\ref{SHBFig} b. where the visible features belong to three different sites 1, 3, and 5 that all share the same field projection. However, for a different case of Figure \ref{SHBFig} c, the visible features show only the ($\Delta$g -$\Delta$e) anti-holes for a pair of Tm$^{3+}$ sites with different magnetic field projections. Thus, with hole spectra from a series of different orientations, the problem then becomes interpreting each spectrum and making an assignment for each feature to a particular site and its corresponding external field projection. This process provides us with a set of splittings that can be used to determine the the tensor values for a given site at a given orientation.

This experiment was conducted in one of the flow cryostats described in Sec.\ref{section3}. The Tm:YGG sample was cooled to 2K and a fixed field of 300mT was applied as the sample was rotated through a series of angles. To create a hole spectrum, the experimental burning sequence at each angle involved a 1ms hole burning pulse, a waiting period of 10 ms for population to decay from the excited state, and finally a  100$\mu$s read pulse with a frequency chirped over a 40MHz region centered at the main hole to reveal any side and anti-hole structure. This sequence used circularly polarized light with $\sim$10mW of power incident on the crystal and was averaged over dozens of scans to produce a single spectrum at each angle. Due to a combination of pump laser linewidth, sample temperature, and limited AOM frequency shifting efficiency, we were only able to monitor and record the frequency change of the first anti-hole, occurring at a frequency offset of ($\Delta_g$ -$\Delta_e$), for all sites during these rotational SHB measurements.

Though not a direct measurement of the tensor values of either the ground or excited state, these values contain information that in conjunction, with the ODNMR data from Section \ref{section5}, is used to compute the spin-Hamiltonian. Tracking these splittings through the series of different angles between an external magnetic field and the symmetric crystal axes, we define a "difference tensor" that best satisfies the measured ($\Delta_g$ -$\Delta_e$) splittings for all sites at all angles.

More specifically, for a specific angle of external field with respect to a particular crystal direction, there is a known projection of the magnetic field on each of the six sites. Our fitting procedure fits $g_{(x,y,z)}$ values in Eq.\ref{splittingEqn} to a particular splitting that is assigned to a site and determines the three associated magnetic field projections ($B_x, B_y, B_z$) at each angle. Thus, one set of fitted values match an entire matrix of magnetic projections for a series of angles with regard to each site to a set of splitting values assigned to each site. The spectral locations of the visible anti-holes at the series of angles compared to the results of this least squares tensor fitting procedure can be seen in Fig.\ref{SHBFig}d. This fitting procedure and analysis were repeated for two differently oriented Tm:YGG crystals, and good agreement was found between the resulting "difference tensors". For more information regarding the fitting procedure, and plots of the fit for the second crystalline direction, consult the supplementary material \cite{Supplementary}.

\section{ODNMR Measurement and Results} \label{section6}

\begin{figure}[t]
\begin{centering}
\includegraphics[width=0.5\textwidth]{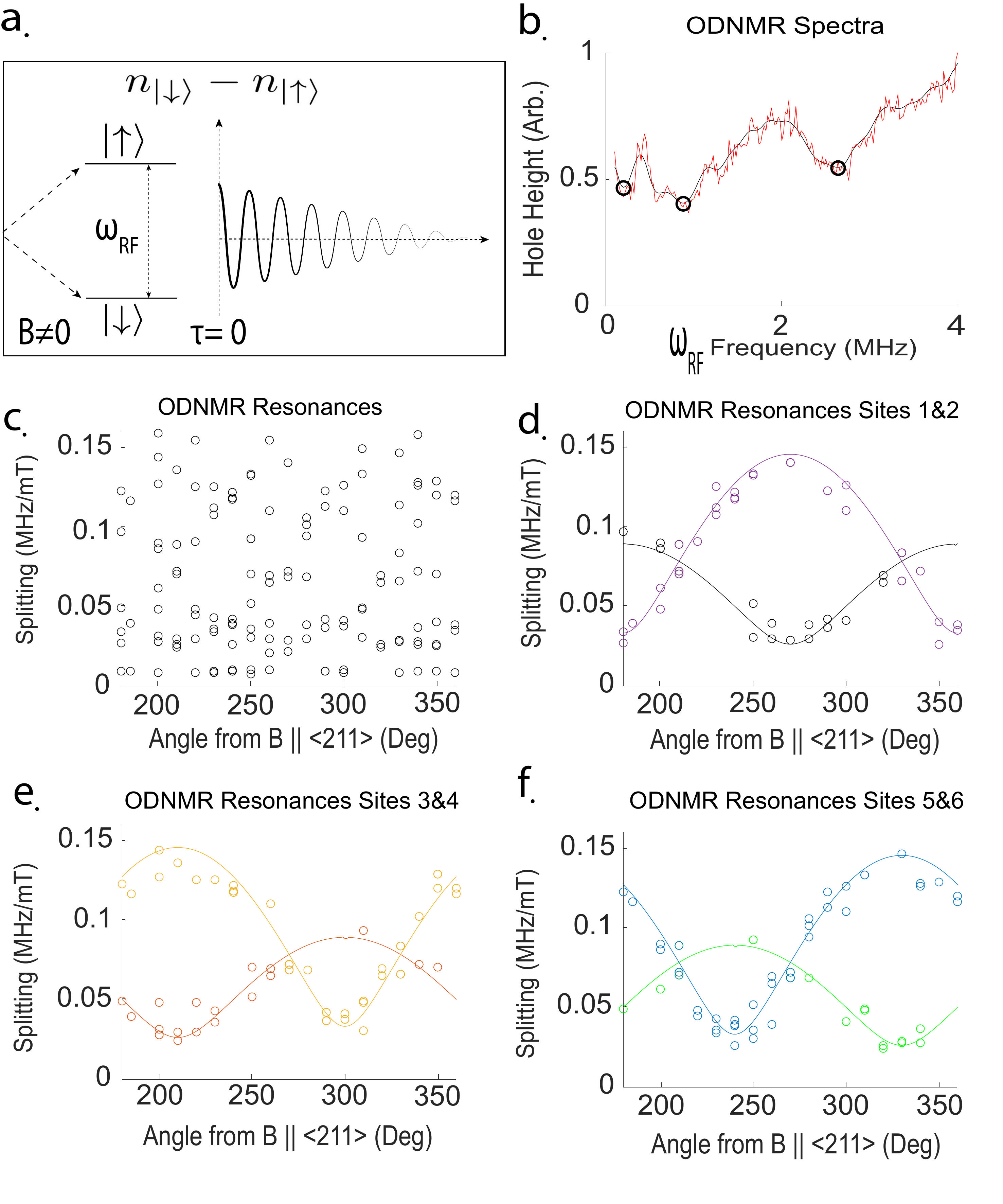}	
\caption {\textbf{a.} Diagrammatic view of the ODNMR experiment, including the driven spin transition that appears at small magnetic fields, and the coherent RF drive that equalizes the population between the $\pm \frac{1}{2}$ spin states. \textbf{b.} The $\omega_{RF}$ frequencies that diminish the height of a spectral hole are identified as resonances by a smoothing and peak finding algorithm and then confirmed manually. A spectrum such as this one is generated for many angles between the field with the crystal $||<211>$ axis  \textbf{c}. A plot of resonances from all spectra at a series of angles between external field  and the crystal $<211>$ direction. \textbf{d,e,f} Site by site fitting of the tensors listed in \ref{gTable} to the resonances from \textbf{c} across a series of angles.}
\label{ODNMRFig}
\end{centering}
\end{figure}

 In addition to SHB, where optical pumping drives changes in hyperfine level population by way of the optical excitation to the excited state followed by spontaneous relaxation, it is also possible to alter the relative hyperfine level populations by directly addressing the microwave transition between them. The relative population between the hyperfine levels can then be observed optically in a double resonance experiment known as optically detected nuclear magnetic resonance, which has been carried out with a number of REICs \cite{Erickson1977,Macfarlane1987,Silversmith1986}. As opposed to SHB, ODNMR provides information directly about either $\Delta_g$  or $\Delta_e$ and thereby supplements the results in the previous section to extract the Hamiltonian in Eq.\ref{Ham}. This approach, shown in Fig. \ref{ODNMRFig}a,  involves first creating a population difference between the hyperfine levels via SHB. While viewing this hole profile optically, an RF pulse is applied that, if resonant with the spin transition at frequency $\Delta_g$, partially re-equalibrates the population, resulting in a change of the absorption depth of the spectral hole. Since the energy difference between hyperfine states is unknown for a given external magnetic field the RF field is swept across a range of frequencies to find the resonance condition $\omega_{RF}=$$\Delta_g$. The excited state splitting $\Delta_e$ can be measured with a similar approach using an altered sequence of optical and RF excitation pulses such that population can be addressed and probed while in the excited state.

This experiment was carried out in one of the helium flow cryostats, equipped with an RF coil on a rotation mount to apply the RF fields. The Tm:YGG sample is cooled to 2K and an external magnetic field of 20-40mT is applied perpendicular to the optical and rotational axis along the crystalline $<$211$>$ direction. Using an altered SHB experimental sequence a spectral hole is prepared and read out, with burning and waiting times adjusted to ensure minimal contribution to the spectral hole depth from atomic population in the excited states while maximizing the repetition rate. Additionally, the duration and relative powers of the burn and read pulses were adjusted to produce a slightly shallower hole that could be read with less noise caused by laser power fluctuations.

 While repeating the hole burning experiment, 100 ms long RF pulses are applied to the crystal. We tuned the frequency of the RF field in 20 kHz steps from 100kHz-4 MHz and five repetitions are averaged to produce spectra as seen in Fig.\ref{ODNMRFig} b. The broadening of the features is due to the convolution of the applied RF field linewidth of 100kHz and the RF resonance in-homogeneous linewidth. To ensure that these features showed the ground-state resonances, we first confirmed that each resonance frequency had a linear shift with increasing external magnetic field, and were absent when no external field was applied.  We optimized the generation of the ODNMR spectra over the parameter space of RF drive power, optical hole burning power, duration, sample temperature, SHB duty cycle, and experimental repetition rate to produce the best signal to noise ratio for these resonance scans.

This experiment was repeated for a series of different angles between the crystal  $<$211$>$ axis and the externally applied magnetic field.  All resonance frequencies were determined with help of a peak finding algorithm, and are displayed in Fig. \ref{ODNMRFig} c. As before, all six Tm sites contribute to this set of resonances. 

Here again, each specific angle of external field with respect to a particular crystalline direction defines six projections of the magnetic field on each site. Each angle is associated with a series of ODNMR resonances as well, such that a vector of ground state tensor values must solve, as closely as possible, a matrix equation that includes magnetic projections on each site for many local angles, to fit over more than 200 equally weighted resonance points. The resulting tensor fits are shown in Fig. \ref{ODNMRFig} d,e,f. Site assignment and fitting was repeated for two differently oriented Tm:YGG crystals that show good agreement with one another. The results of the second fit can be seen in the supplementary material \cite{Supplementary}. It is interesting to note that the calculated tensor values are comparable to those measured with TmGG \cite{Jones1968}. This gives us additional confidence in the results, as in the case of the confirmation between Tm:YAG and TmAG \cite{Louchet2007,Schmidt1978}.

This fitting procedure produces the three ground state tensor values $\Lambda_{g,(x,y,z)}$, and in conjunction with the ``difference tensor" fits from the previous section, $\Lambda_{e,(x,y,z)}$ can also be determined. These results for the six spin-Hamiltonian components averaged over the four directional fits of sections \ref{section5} \& \ref{section6} are shown in Table \ref{gTable}, where the signs are given by convention.  Unfortunately, both spectral hole burning and ODNMR yield only the magnitude but not the sign of the different parameters because we cannot differentiate one hyperfine state from another. Nonetheless, and similarly to Ref.[28]\cite{McAuslan2012}, for our purposes it is often enough to know the relative energy difference between the two hyperfine states of each electronic level, and not whether it happens to be the spin up or spin down state since their energy shifts will be symmetric. 

\begin{center}
\begin{table}
\begin{tabular}{ |c|c|c|c| } 
\hline
 & $^3$H$_6$(g) & $^3$H$_4$(e) \\
\hline
$g_n\beta_n$ & -3.53(MHz/T) &  -3.53(MHz/T) \cite{Schmidt1978}\\
$g_j$ & 1.16 & 0.8 \\  
A$_J$ & -470.3(MHz) & -678.3(MHz) \\ 						      %\multirow{3}{4em}{Multiple row}
A$_J\Lambda_{x}$ & -7.23 $\times 10^{-4}$ & -1.55 $ \times 10^{-4}$ \\ % Unitless values for now.
A$_J\Lambda_{y}$ & -4.47 $\times 10^{-3}$ & -3.95 $ \times 10^{-3}$ \\ % Unitless values for now.
A$_J\Lambda_{z}$ & -9.99 $\times 10^{-4}$ & -5.57 $ \times 10^{-4}$ \\ % Unitless values for now.
$g_x$  & 27 $\pm$ 1.7 (MHz/T) &  7 $\pm$ 2.5 (MHz/T)\\ 
$g_y$ & 146 $\pm$ 1.5(MHz/T) &  92 $\pm$ 2.8 (MHz/T) \\ 
$g_z$ & 36 $\pm$ 2.6 (MHz/T) &  16 $\pm$ 3.1 (MHz/T) \\ 
\hline
\end{tabular}
\caption{Spin Hamiltonian values. The $g_j$ values are calculated using the general formalism described in Ref.[14]\cite{Abragam2012}. The A$_J$ values are from Ref.[7]\cite{Guillot-Noel2005} and can be recalculated from Ref.[14]\cite{Abragam2012} as they depend only on the expectation value for electron radius for this Tm$^{3+}$ transition. The $g_{x,y,z}$ values for the excited and ground states are measured in this work. Together with the listed constants, they allow calculating all $\Lambda_{J,\alpha}$ values using Eq.\ref{gvalEqn}.}
 \label{gTable}
\end{table}
\end{center}

\section{Verification of results} \label{section7}
In the following, we detail additional measurements based on alternative methods and on different Tm:YGG crystals. While being performed with a reduced set of parameters that does not allow independently establishing the Hamiltonian in Eq. \ref{Ham}, they nevertheless allow verifying the results presented in section \ref{section6}.

\begin{figure}[t]
\begin{centering}
\includegraphics[width=0.5\textwidth]{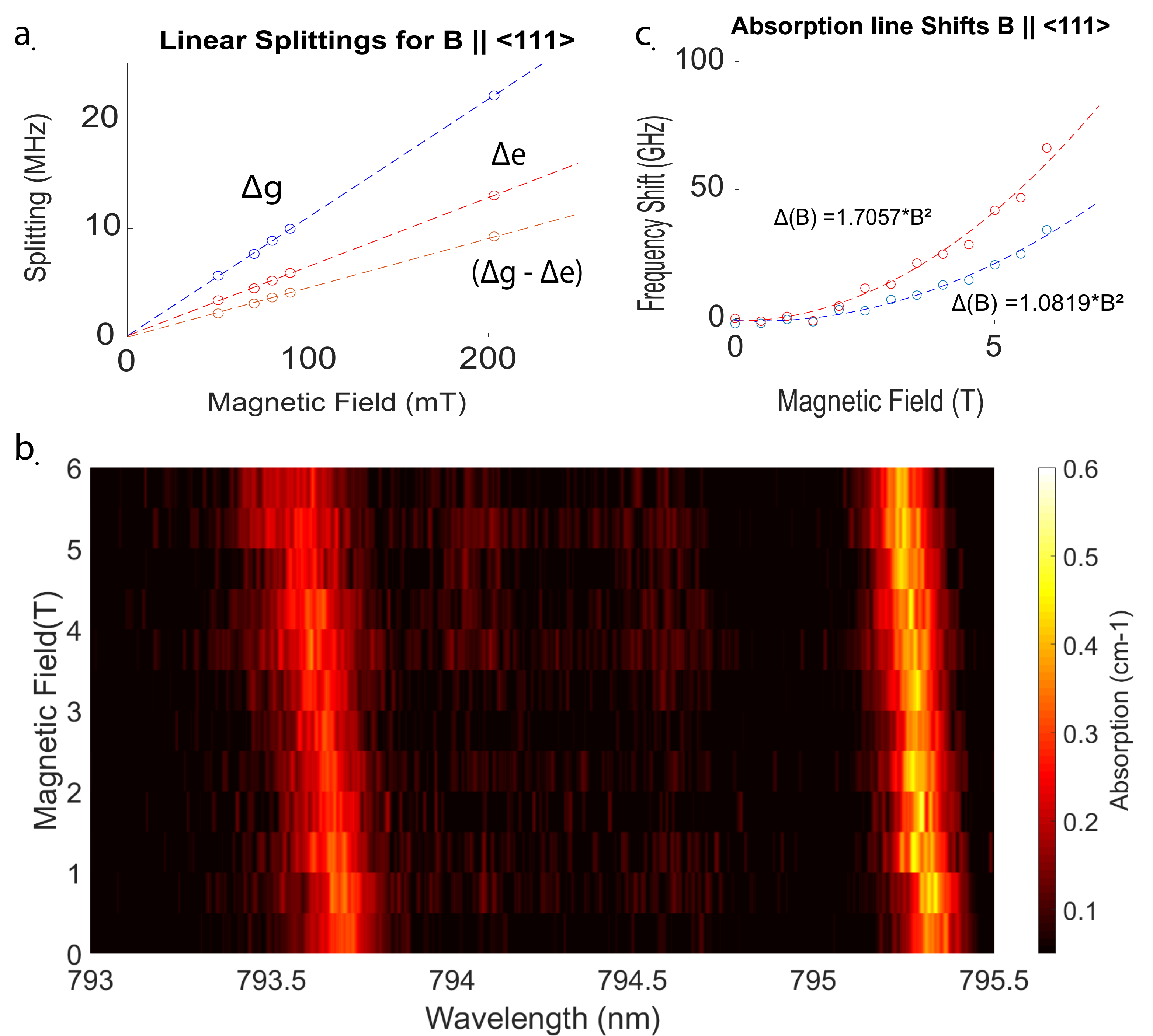}    
\caption {\textbf{a.} The linear splitting of the anti-holes shown in Fig.\ref{SHBFig} b. for the B $||$ $<$111$>$ orientation yields values of 45 and 108$\pm$ 2 MHz/T and a side hole splitting of 63$\pm$ 2 MHz/T. \textbf{b.} Absolute absorption frequency for the optical transitions from the lowest ground state crystal field level to the first (795.325 nm) and second(793.7nm) excited state crystal field levels with increasing magnetic field. \textbf{c.} A closer look at the behavior of these transitions reveals the expected quadratic nature of the frequency shift.}
\label{ConfirmsFig}
\end{centering}
\end{figure}

\subsection{Linear Zeeman shift}
The enhanced effective Zeeman interaction given by the first two terms of Eq.(\ref{Ham}) results in a linear splitting of the hyperfine levels in both the ground and excited states.  As discussed in section \ref{section4}, the Zeeman coefficient can be measured by first applying a magnetic field, then exposing the sample to narrow-band laser light, and finally probing the resulting sequence of holes and anti-holes by frequency scanning a weak laser across the created spectral features.

This experiment was performed inside the ADR and at a temperature of 500 mK. In this case, the magnetic field could only be applied along the crystal's $<$111$>$ direction, resulting in magnetic projections of equal magnitude on the Y and Z components of sites 1,3, and 5, and the X and Z components of sites 2,4, and 6. The holeburning laser was frequency locked to around 10 kHz linewidth by means of a stable optical reference cavity and the Pound-Drever-Hall method, and we employed 1 mW of power incident on the crystal, burning pulses of 200ms duration, 10ms of waiting time to avoid spontaneously emitted signal, and a 1ms long read pulse. The lower temperature and narrower laser linewidth (compared to the measurements described before) allowed resolving all expected SHB features, including a pair of anti holes and a side hole, as seen in Fig. \ref{SHBFig} b. for a field of 90 mT.

The splitting of these spectral features as a function of magnetic field shown in Fig.\ref{ConfirmsFig} a. confirms the linear nature of the Zeeman interaction with splittings of the ground and excited states of 108 $\pm$ 2 MHz/T and 63 $\pm$ 2 MHz/T, respectively. This is in good agreement with the values calculated from the measured tensors, which we found to be $\Delta_g = 106 \pm 2.2$ MHz/T, $\Delta_e = 66 \pm 1.3$ MHz/T for sites 1,3, and 5 at this orientation.

\subsection{Quadratic Zeeman shift}

The quadratic Zeeman interaction given by the third term of Eq.(\ref{Ham}) shifts each electronic crystal field level, characterized by the total angular momentum J, parabolically due to the so-called VanVleck paramagnetism \cite{VanVleck1932}.   Each crystal field level shifts with its own rate, as depicted in Fig.\ref{SHBFig} a. and described by its hyperfine tensor. This creates an overall shift of the optical transition frequency with a $B^2$ dependence that varies based on the external field direction. It can be calculated using the ground-state and excited-state tensor values given in Table \ref{gTable}.

To measure the quadratic Zeeman effect, we used a flow cryostat operating at 5K with an applied magnetic field along the $<$111$>$ direction of the Tm:YGG crystal.  We performed white light absorption spectroscopy, similar to Ref.[16]\cite{Veissier2016}, i.e. we directed collimated CW white light through the crystal and to an optical spectrum analyzer to observe the entire inhomogeneous line of the $^3$H$_6 \Leftrightarrow ^3$H$_4$ transition at 795.325nm (see Fig.\ref{ConfirmsFig}b). The optical transition from lowest crystal field level of the ground state to the second crystal field level of the excited state at 793.7 nm was visible as well, confirming the expected crystal field spacing of 26 $cm^{-1}$ in Ref.[3]\cite{Sun2005}. We monitored the absorption frequencies for this pair of transitions as the external magnetic field was increased from 0 to 6 T. The results are depicted in Fig. \ref{ConfirmsFig} c. A quadratic fit yields a frequency shift of $\Delta=1.08 \pm 0.24$  GHz/T$^2$, which is in good agreement with the value of  $\Delta_p=1.09 \pm 0.25$ GHz/T$^2$ that we calculated from the measured hyperfine tensors. %% Re-check the error bars. 

\section{Optical Clock Transitions and Special Directions} \label{section8}

The frequency of the transition between two hyperfine states--one belonging to the electronic ground state and one of the excited state--depends on the applied magnetic field through the linear and quadratic Zeeman effects. Small field fluctuations therefore cause spectral diffusion and hence a reduction of the optical coherence time. It is sometimes possible to choose a magnetic field magnitude and a crystal orientation for which the transition energy becomes, to first order, insensitive to fluctuations in the magnetic field. These transitions are referred to as optical clock transition and are analogous to the ZEFOZ transitions from Ref.[5,28,30,31] \cite{Fraval2004,Longdell2006,Loveric2011,McAuslan2012}. Clock transitions are extremely powerful tools used for reducing decoherence due to noisy magnetic fields \cite{Kielpinski1013}. These clock transitions exist in Tm:YGG and can be found for a particular site using the procedure detailed below. 

Inspecting the third term of Eq. \ref{Ham} and taking into account that the elements of the hyperfine tensor $\Gamma_J$ are larger in the ground state than in the excited state (see Ref.[16]\cite{Veissier2016} for a detailed explanation in Tm:YAG) the optical transition energy change caused by the quadratic Zeeman effect creates an increase in transition energy with magnetic field. This is depicted theoretically in Fig.\ref{SHBFig} a. and confirmed from the data in Fig. \ref{ConfirmsFig}b.

\begin{figure}[t]
\begin{centering}
\includegraphics[width=0.5\textwidth]{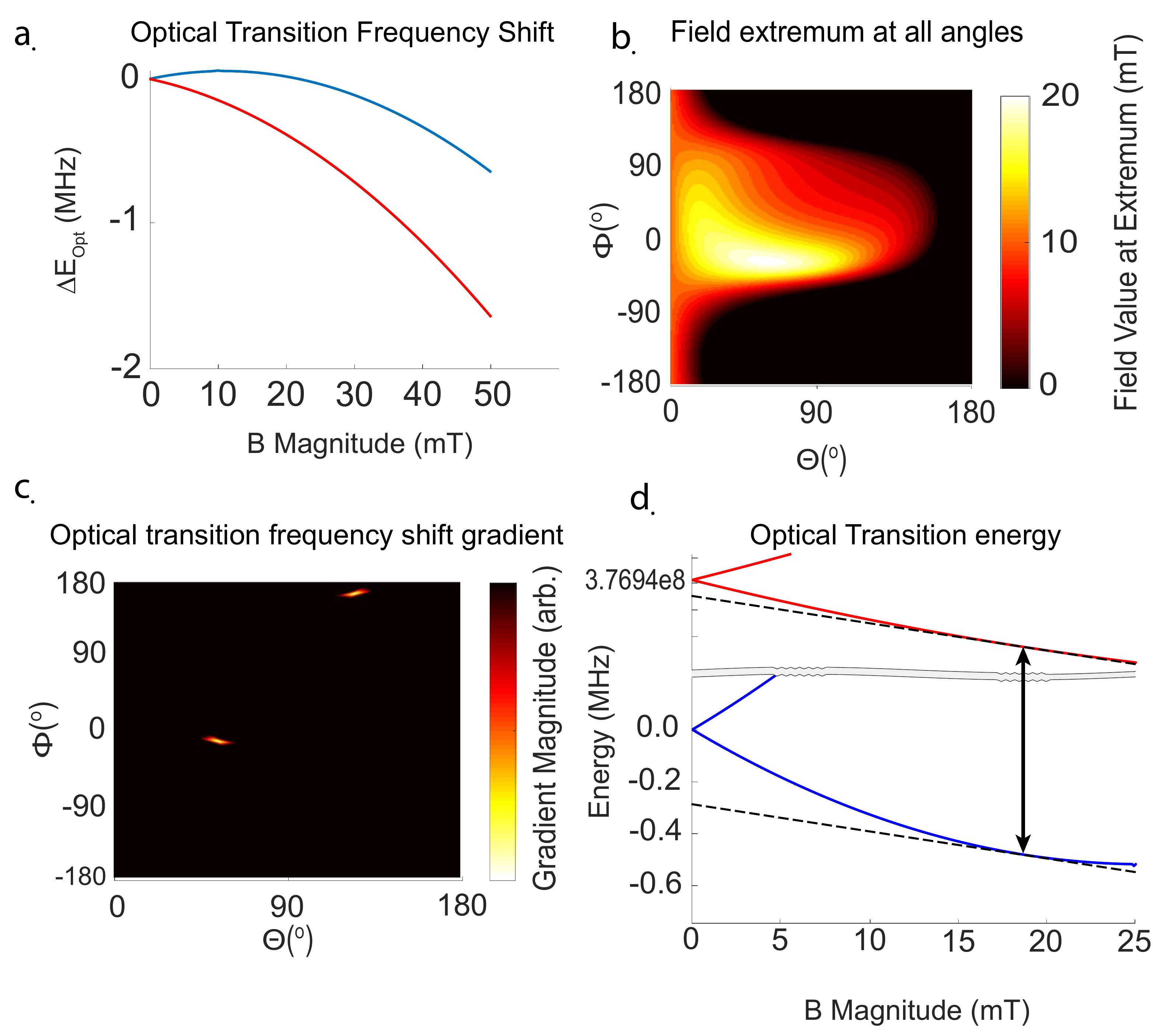}	
\caption {\textbf{a.} The shift in the optical transition frequency follows a quadratic dependence with an extrema that, for every orientation, appears at a positive field for the spin conserving -1/2  $\Leftrightarrow$ -1/2  transition (blue) and at some negative field for the spin conserving +1/2  $\Leftrightarrow$ +1/2 transitions (red). \textbf{b.} For the -1/2  $\Leftrightarrow$ -1/2 spin conserving optical transition and for all angles, the magnetic field value of this extrema is identified. \textbf{c.} Angular gradient of the optical transition frequency at a field of 19mT. Two points are visible for which the angular gradient approaches zero. \textbf{d.} Energy shifts of the two -1/2 spin states at this particular direction vs. magnetic field magnitude. At 19mT, the slope of the energy change for both states with change in field magnitude is equal (along with the other derivatives) and an optical clock transition occurs.}
\label{ZEFOZ}
\end{centering}
\end{figure}

Furthermore, looking at the first and second term of Eq. \ref{Ham}, which describe the linear Zeeman shift, we find that there exist ground and excited state pairs of hyperfine states for which the optical transition frequency decreases with applied magnetic field. Hence, for certain combinations of crystal orientation (or rather orientation of a local Tm$^{3+}$ site) with respect to the magnetic field, magnetic field strength and spin values, the linear and quadratic Zeeman effects oppose each other, and an optical clock transition can be observed \cite{Tongning2015}.  This transition must have a field magnitude insensitive point, which is shown in Fig. \ref{ZEFOZ} a as the shift in optical transition frequency $\Delta E_{opt}(|B|,\theta,\phi)$ vs. applied field magnitude.

In these cases, the spin-Hamiltonian terms form a quadratic function that must have an extremum, i.e. a field-magnitude-insensitive point,  at some positive field magnitude. The first condition of an optical clock transition, a field magnitude $|\boldsymbol{B}|$ such that $\frac{\partial \Delta E_{opt}}{\partial \boldsymbol{|\vec{B}|}}=0$, is calculated for every orientation of magnetic field relative to the "Site 1" spin -1/2 conserving optical transition in Fig. \ref{ZEFOZ} b using a mesh size of 1mT and 1 degree angular steps.

For the remaining angular derivatives ($\frac{\partial \Delta E_{opt}}{\partial B_{\theta}},\frac{\partial \Delta E_{opt}}{\partial B_{\phi}}=0$), there must also be angles for which $\Delta E_{opt}$ is also insensitive to first order angular change to complete each optical clock transition. Though all orientations have some field magnitude that results in an invariant point, only some orientations have angular gradients that simultaneously go to zero. Plotted in Fig. \ref{ZEFOZ}c is the magnitude of the angular gradient for both spin conserving optical transitions vs the optical transition frequency for all angles of magnetic field with respect to the crystal axes. Two angular points stand out, one for each spin conserving transition, whose gradient value becomes exceedingly small. By cross referencing the points of small angular gradient with the field at which the magnitude extrema appears, we can find an orientation that simultaneously brings all three derivatives to zero. In the case of the spin -1/2 conserving optical transition for "Site 1", we arrive at a set of coordinates $\boldsymbol{\vec{B}_{|B|,\theta,\phi}}= (19mT, 55^o,-15^o)$ that correspond to an optical clock transition. The invariant energy levels of this transition at this field are shown in Fig. \ref{ZEFOZ}d.  This procedure can be repeated for the remaining sites and potential optical transitions (spin crossing or preserving), giving us the list of optical clock transitions in this material shown in Table \ref{ZEFOZ Table}. 

Past studies characterize the benefit to coherence gained by using a field invariant direction with the definition of transition curvature given by Ref.[30] \cite{Longdell2006}. All transitions in Table \ref{ZEFOZ Table} share the same curvature of $\sim 36 $Hz$/G^2$.  Other REIC samples previously brought to invariant points with this magnitude of curvature have resulted in orders of magnitude improvement to spin coherence times. However, the amount of improvement in the case of Tm:YGG will depend heavily on the decoherence mechanisms involved for this optical transition.

\begin{center}
\begin{table}
\begin{tabular}{ |c|c|c|c|c| } 
\hline
  Site \# & $|\boldsymbol{B}|$ & $\boldsymbol{\vec{B}_{\theta}}$ & $\boldsymbol{\vec{B}_{\phi}} $ & Spin Level Transition    \\
\hline
1 & 19 mT &  55$^o$ & -15$^o$  & -1/2 $\Leftrightarrow$  -1/2 \\
1 & 19 mT &  125$^o$ & 166$^o$  &+1/2 $\Leftrightarrow$ +1/2\\
1 & 36 mT &  64$^o$ & -150$^o$  &+1/2 $\Leftrightarrow$ -1/2\\
1 & 36 mT &  117$^o$ & 31$^o$  &-1/2 $\Leftrightarrow$ +1/2\\
\hline
2 & 19 mT &  54$^o$ & 76$^o$  & -1/2 $\Leftrightarrow$  -1/2\\
2 & 19 mT &  125$^o$ & -105$^o$  & +1/2 $\Leftrightarrow$ +1/2\\
2 & 36 mT &  116$^o$ & 120$^o$  & -1/2 $\Leftrightarrow$ +1/2\\
2 & 36 mT &  63$^o$ & -60$^o$  & +1/2 $\Leftrightarrow$ -1/2\\
\hline
3 & 19 mT &  102$^o$ & 54$^o$  &-1/2 $\Leftrightarrow$  -1/2\\
3 & 19 mT &  79$^o$ & -127$^o$  &+1/2 $\Leftrightarrow$ +1/2\\
3 & 36 mT &  64$^o$ & 120$^o$  &+1/2 $\Leftrightarrow$  -1/2\\
3 & 36 mT &  118$^o$ & 60$^o$  &-1/2 $\Leftrightarrow$  +1/2\\
\hline
4 & 19 mT &  38$^o$ & 20$^o$  &-1/2 $\Leftrightarrow$  -1/2\\
4 & 19 mT &  143$^o$ & -160$^o$  &+1/2 $\Leftrightarrow$  +1/2\\
4 & 36 mT &  140$^o$ & -45$^o$  &-1/2 $\Leftrightarrow$  +1/2\\
4 & 36 mT &  40$^o$ & 135$^o$  &+1/2 $\Leftrightarrow$  -1/2\\
\hline
5 & 19 mT &  38$^o$ & 110$^o$  &-1/2  $\Leftrightarrow$  -1/2\\
5 & 19 mT &  142$^o$ & -70$^o$  &+1/2  $\Leftrightarrow$  +1/2\\
5 & 36 mT &  140$^o$ & 45$^o$  &-1/2 $\Leftrightarrow$  +1/2\\
5 & 19 mT &  40$^o$ & -135$^o$  &+1/2  $\Leftrightarrow$  -1/2\\
\hline
6 & 19 mT &  78$^o$ & 37$^o$  &-1/2 $\Leftrightarrow$  -1/2\\
6 & 19 mT &  102$^o$ & -144$^o$  &+1/2 $\Leftrightarrow$  +1/2\\
6 & 36 mT &  62$^o$ & 30$^o$  &-1/2 $\Leftrightarrow$  +1/2\\
6 & 36 mT &  118$^o$ & -150$^o$  &+1/2 $\Leftrightarrow$  -1/2\\
\hline
\end{tabular}
\caption{Optical clock transitions with angular coordinates of $\boldsymbol{\vec{B}}$ given relative to the ($<$100$>$,$<$010$>$,$<$001$>$) crystalline axes. These positive and negative projections are based on the unit cell directions for the 6 sites for the Tm:YGG.}
 \label{ZEFOZ Table}
\end{table}
\end{center}

Though the clock transition directions are important for maximizing optical coherence time, alternative crystal properties can also be optimized by choosing other external field directions.  Following the arguments from Ref.[34]\cite{Pascual-Winter2012}, we have found other orientations that are capable of minimizing the spin inhomogeneous broadening for the ground state spin transition. Similarly, according to Ref.[22,35]\cite{Louchet2007,deSeze2006} another orientation exists that allows maximizing the branching ratio of the available lambda system in this material. Details on the calculation of these orientations are available in the supplementary material\cite{Supplementary}.

\section{Conclusion} \label{section9}

We conducted a series of spectroscopic studies that allowed us to measure the full spin-Hamiltonian for Tm:YGG. Despite the multiple sites, a combination of SHB and ODNMR determined both the ground-state and excited-state hyperfine tensors. These tensors were confirmed to match the results of a pair of independent experiments with magnetic field directions limited to the $<$111$>$ crystal axis. This new knowledge allowed us to determine a series of directions that should create optical clock transitions for specific thulium sites, and in turn, enhance the optical coherence of the transition in this material. This confirms the potential of Tm:YGG crystals for quantum networking and quantum memory applications.

\subsection*{Acknowledgments}  
All research at Montana State University was sponsored by Air Force Research Laboratory under agreement number FA8750-20-1-1009. Additionally, we acknowledge funding through the Netherlands Organization for Scientific Research (NWO), and the European Union’s Horizon 2020 research and innovation program under grant agreement No 820445 and project name Quantum Internet Alliance. Finally, we appreciate the help of Thomas Rust for his work cutting and polishing some of the many crystal samples used in this work.

%\section*{References}
%\bibliographystyle{ieeetr}
%\bibliography{YGGOrientBib}	

\end{document}

\captionsetup[figure]{labelfont={bf},name={SIFig.},labelsep=period}

	\title{Measurement of the Thulium Ion Spin Hamiltonian Within a Yttrium Gallium Garnet Host Crystal}
	
	\author{Jacob H. Davidson}
	\affiliation{Qutech and Kavli Institute of Nanoscience, Delft University of Technology, Delft, The Netherlands}
	%-----------------------------
	\author{Philip J.T. Woodburn}
	\affiliation{Department of Physics, Montana State University, Bozeman, Montana 59717, USA}
	%-----------------------------
	\author{Aaron D. Marsh}
	\affiliation{Department of Physics, Montana State University, Bozeman, Montana 59717, USA}
	%-----------------------------
	\author{Kyle J. Olson}
	\affiliation{Department of Physics, Montana State University, Bozeman, Montana 59717, USA}
	%-----------------------------
	\author{Adam Olivera}
	\affiliation{Department of Physics, Montana State University, Bozeman, Montana 59717, USA}
	%-----------------------------
	\author{Antariksha Das}
	\affiliation{Qutech and Kavli Institute of Nanoscience, Delft University of Technology, Delft, The Netherlands}
	%-----------------------------
	\author{Mohsen Falamarzi Askarani}
	\affiliation{Qutech and Kavli Institute of Nanoscience, Delft University of Technology, Delft, The Netherlands}
	%-----------------------------
	\author{Wolfgang Tittel}
	\affiliation{Qutech and Kavli Institute of Nanoscience, Delft University of Technology, Delft, The Netherlands} 
	%-----------------------------
	\author{Rufus L. Cone}
	\affiliation{Department of Physics, Montana State University, Bozeman, Montana 59717, USA}
	%-----------------------------
	\author{Charles W. Thiel}
	\affiliation{Department of Physics, Montana State University, Bozeman, Montana 59717, USA}
	
	%\date{\today}
	
	%-----------------------------	

 \maketitle

\section{Fitting details}

To fit the data present in each of the angular scans we begin with a simple linear regression problem of the form:
\begin{equation}
 \boldsymbol{A}\vec{\lambda}=\vec{\Delta}
\label{matrixeq1}
\end{equation}
where  $\boldsymbol{A}$ is a m x 3 matrix of magnetic field projections, where each of the m rows, representing a particular ion site at a particular angle, is given by $B_x, B_y, B_z$. The $\vec{\Delta}$ is a vector that contains all the measured splittings from either ODNMR or SHB given the particular fit. Finally, the $\vec{\lambda}$ vector contains the hyperfine tensor values that solve this system of equations for the series of ion sites at a series of angles.  Since no tensor values will fit the measured splittings with zero error and $\boldsymbol{A}$ may not be invertable, the then the matrix equation in Eq. \ref{matrixeq1} must be generalized to its ordinary least squares version given by:
\begin{equation}
\vec{\lambda}=\big(\boldsymbol{A^T}\boldsymbol{A}\big)^{-1}\boldsymbol{A^T}\vec{\Delta}
\label{matrixeq2}
\end{equation}
which gives the closest possible fit for the linear sytem of equations. Generalizing again, since the splittings do not depend linearly on the magnetic field projections, we solve the new matrix equation using the nonlinear modeling function of matlab for which the user supplies, the input vectors, $\vec{\lambda}$, and $\vec{\Delta}$, and the nonlinear model that describes their relationship, in this case,
\begin{equation}
\Delta_g=\sqrt {g_{x}^2B_x^2+ g_{y}^2B_y^2+ g_{z}^2B_z^2}
\end{equation}
or
\begin{equation}
\Delta_g-\Delta_e=\sqrt {g_{x}^2B_x^2+ g_{y}^2B_y^2+ g_{z}^2B_z^2}- \sqrt {e_{x}^2B_x^2+ e_{y}^2B_y^2+ e_{z}^2B_z^2}
\end{equation}
for the ground and ground minus excited state splittings respectively. 
The only task that remaining is to make sure that each row of projections is associated with the correct resonance splitting to produce the final set of values. These associations were made manually by relating each resonance to the most likely sites and angles using similar initial measurements from \cite{Jones1968}.

All magnetic field projections are computed based on a unit vector magnetic field in relation to the site axes pictured in main text Figure 1. Each sites' axes given in the crystalline cubic cell frame are assumed to be:
\begin{center}
\begin{table}
\begin{tabular}{ |c|c|c|c| } 
\hline
Site  & $\hat{x}$ & $\hat{y}$ & $\hat{z}$ \\
\hline
 1 (Black)  & (1 -1 0) & (1 1 0) & (0 0 1)\\
 2 (Purple) & (1 1 0) & (-1 1 0) & (0 0 1)\\
 3 (Orange) & (0 1 -1) & (0 1 1) & (1 0 0)\\
 4 (Yellow) & (0 1 1) & (0 -1 1) & (1 0 0)\\
 5 (Green) & (-1 0 1) & (1 0 1) & (0 1 0)\\
 6 (Black) & (1 0 1) & (1 0 -1) & (0 1 0)\\
\hline
\end{tabular}
\caption{Defined directions in the crystal. These directions were normalized to compute all projections of the magnetic fields. The positive and negative axes are assumed by convention to keep the site $\hat{z}$ directions pointing into the unit cell. }
 \label{AxisTable}
\end{table}
\end{center}
Inverting the sign assumptions present in these directional definitions can result in some underlying changes to the signs but not the magnitudes of the determined tensor values and special directions relative to each defined axis \cite{McAuslan2012}. 

From the set of all the points for a particular measurement, resonance points which showed no angular dependence or extremely weak signatures were also omitted from the fits. Many of the omitted resonance points appeared at all angles of ODNMR with a splitting of 9-11 MHz/T and are assumed to be the gallium nuclear spin splittings of the host crystal. Finally, a fixed angular correction was made to all points for each particular set of data to account for the experimental uncertainty in the crystal orientation.

\section{Secondary Fit of the SHB Difference Tensor}
\begin{figure}[t]
 \centering
\includegraphics[width=0.5\textwidth]{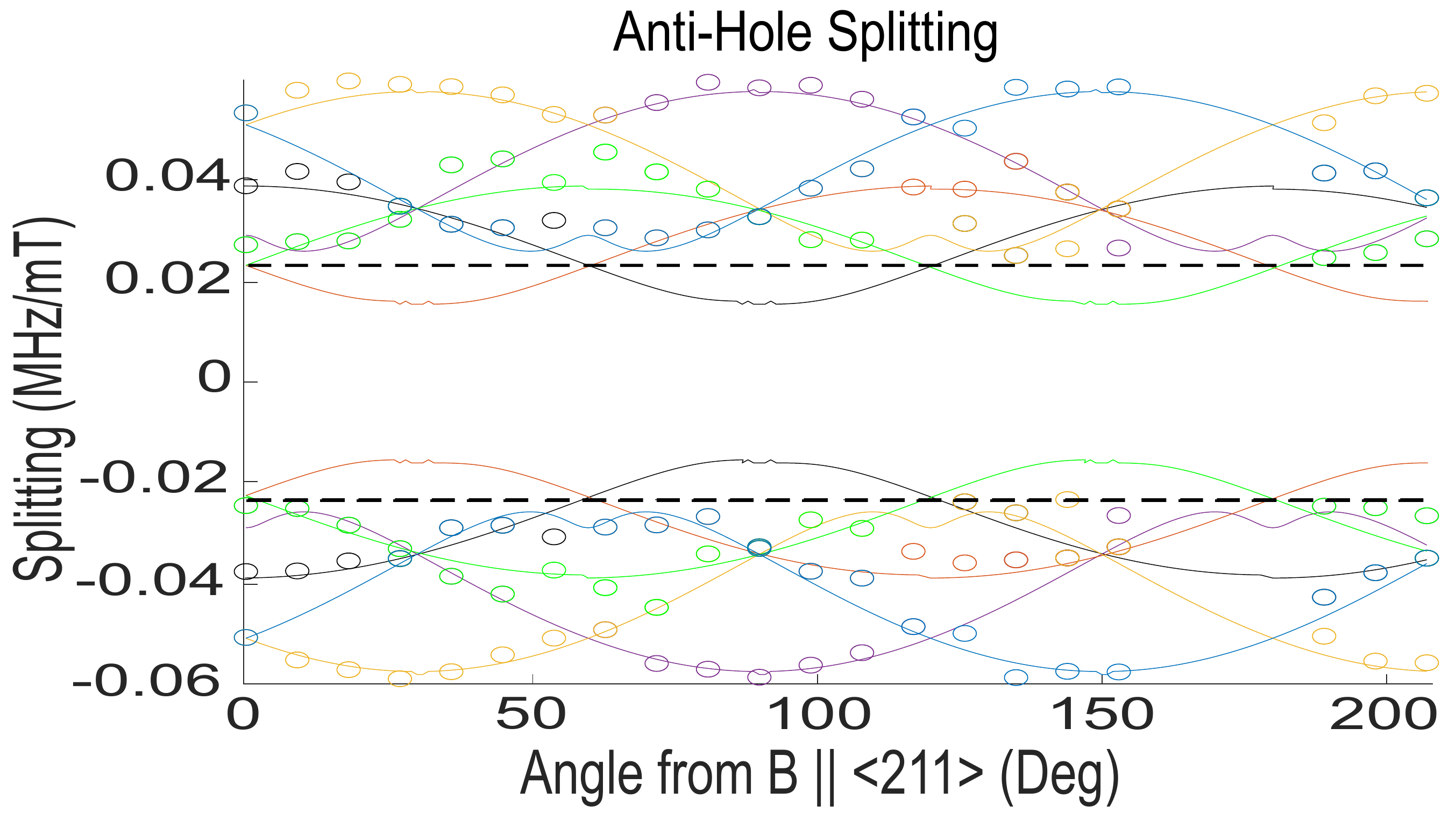}	
\caption {Secondary SHB directional fit. \textbf{a.} Spectral hole burning anti hole splittings measured for an optical axis along the crystalline $<$110$>$ axis. Dotted line indicate the minimum measurable splitting due to the width of the central spectral hole.}
\label{SIFig1}
\end{figure}

Given the experimental uncertainty in angles and orientation, and the manually conducted site assignment for each of the splittings it is important to verify the tensor values that are produced by our fitting procedure explain the splitting behavior for multiple crystalline directions. To account for this we rotated the crystal to a new optical axis, defining a new plane in which to rotate the B field, and repeated our angular rotation while collecting SHB data for the new direction. The fitting procedure described above was applied to the resonances of the new direction independently of the data from previous directions. The resulting tensor values, with fits pictured in Fig.\ref{SIFig1}, show close agreement with the values produced by the fit of the first direction.  The tensor values from both directions are averaged together to give the final values listed for the excited state splittings in the main text. 

\section{Secondary Fit of the ODNMR directional Tensor}
\begin{figure}[t]
 \centering
\includegraphics[width=0.5\textwidth]{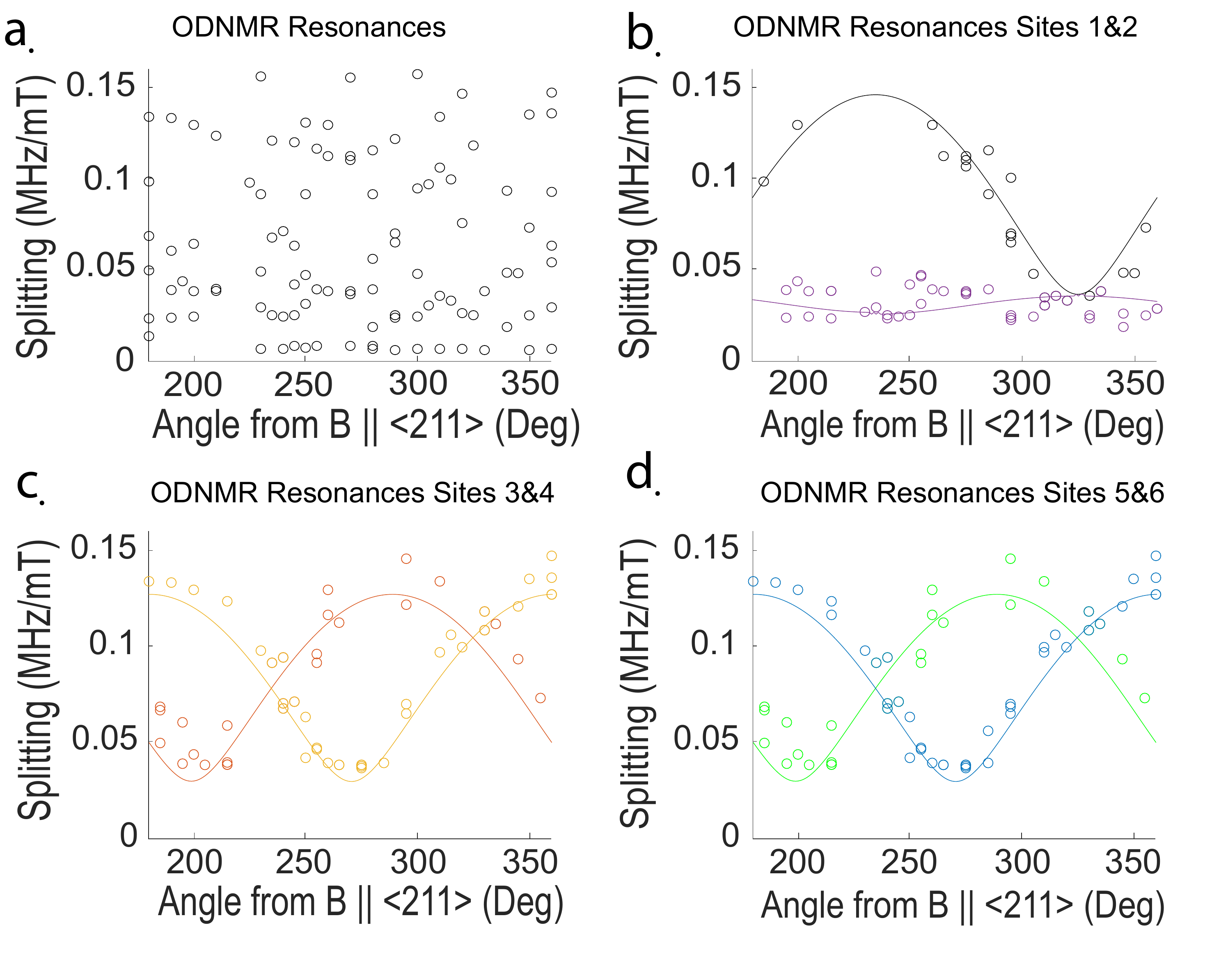}	
\caption {Secondary ODNMR directional fit using ODNMR data. \textbf{a.} All ODNMR resonances measured for an optical axis along the crystalline $<$110$>$ axis.  \textbf{b,c,d.} Site by site fitting of the tensors listed in the main text to the resonances from \textbf{a} across a series of angles.}
\label{SIFig2}
\end{figure}

Following the same logic for all rotational measurements, we conducted ODNMR measurements while rotating around a pair of different optical axes as well. The resulting tensor values, with fits pictured in Fig.\ref{SIFig2}, also show good agreement with the values produced by the fit of the first direction.  The tensor values from both directions are averaged together to give the final values listed for the ground state splittings in the main text. 

\section{Minimize the Spin-Inhomogeneous Broadening}
In the main text, the results for an optical clock transition are depicted to show an insensitive point between a pair of energy levels that belong to the optical transition for the $Tm^{3+}$ ions in this material. The purpose of this point is to minimize the contribution of magnetic fluctuations to the uncertainty in optical transition energy. A similar analogy can be drawn to the pair of ground state hyperfine levels if that is instead the transition of interest. Since the nuclear spin of thulium is $1/2$ however, these levels will always experience a linear splitting relative to the magnitude of the applied magnetic field. In other words, a complete ZEFOZ point is not accessible in this system. However, by choosing a specific orientation in the hyperfine tensor space, as in the case of \cite{Pascual-Winter2012}, the effects on these spins due to orientation fluctuations of the local magnetic field can be minimized. 

\begin{figure}[t]
 \centering
\includegraphics[width=\textwidth]{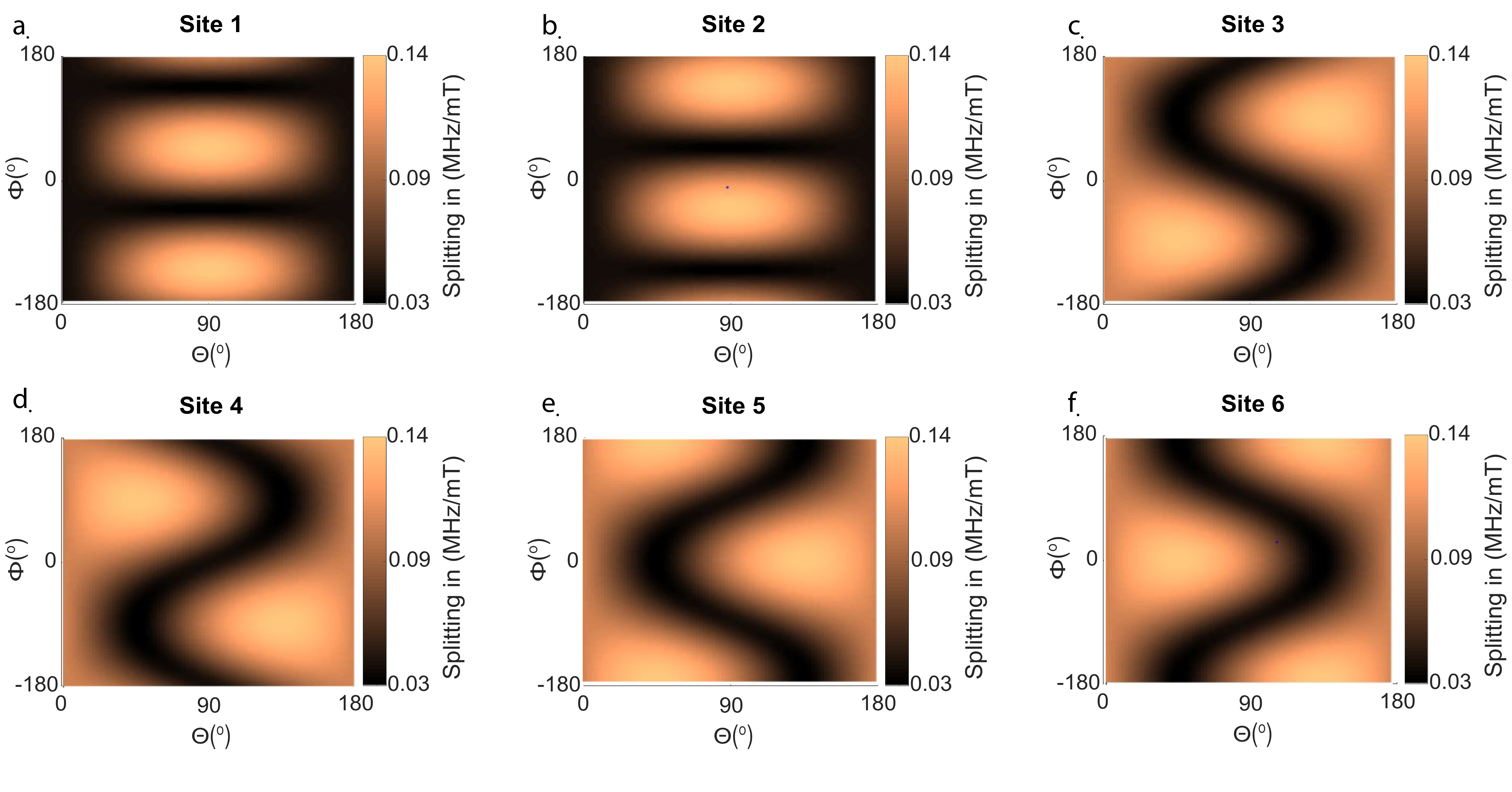}	
\caption {\textbf{a-f.}Site by site orientation map of the ground state splittings. The highlighted extrema of each surface indicate the orientations of improved sensitivity to angular magnetic field fluctuations.}
\label{SIFig3}
\end{figure}

To improve future possibilities for spin control in Tm:YGG a crystal with similar spin coherence to Tm:YAG, but missing the complex cross relaxation mechanism caused by aluminum \cite{Ahlefeldt2015}, we have conducted the same analysis as in the case of \cite{Pascual-Winter2012}. We compute for some magnetic field amplitude, the magnitude at all angles of the splitting of the ground state spin levels. Finding for each of the present sites, extrema in this angular splitting landscape, we pinpoint orientations where the splitting between hyperfine states has the smallest variation with respect to local angular change of an external magnetic field. The results of this computation are shown for each of the six sites in Fig.\ref{SIFig3}.
These external field orientations, computed from the tensors, will result in the narrowest possible spin transition in-homogeneous broadening in this material, and have potential to enable dynamical decoupling and spin control sequences.

\section{Branching ratio simulations and directions}

An additional aspect of interest in these rare earth ion-doped crystal systems is to create a lambda system where two lower energy levels are each connected via optical transition to a single excited state \cite{Louchet2007,deSeze2006,Guillot-Noel2005}. Such a system is relevant to a number of different quantum memory schemes. Since there are only a pair of degenerate hyperfine levels in this spin 1/2 system, the pair of ground states that couple to a single excited level must be the $\pm 1/2$ ground-state hyperfine levels. According to selection rules, optical transitions should not be capable of creating nuclear spin flips making a lambda system with these couplings potentially impossible. However, due to the crystal field mixing of the electronic and nuclear wave-functions such spin crossing transitions become weakly allowed, alongside their spin preserving counterparts. By describing an effective magnetic field felt by the nuclear spin due to each of the nearest crystal field levels,  the amount of spin mixing that is possible can be calculated from the hyperfine tensor values of Tm:YGG. Following the definition of an effective B field vector in the ground and excited states from\cite{deSeze2006}, we compute the likely branching ratio for each ion site at all possible orientations to determine which angles may yield the best spin crossing transition probability for each site in Tm:YGG.
 \begin{figure}[t]
 \centering
\includegraphics[width=\textwidth]{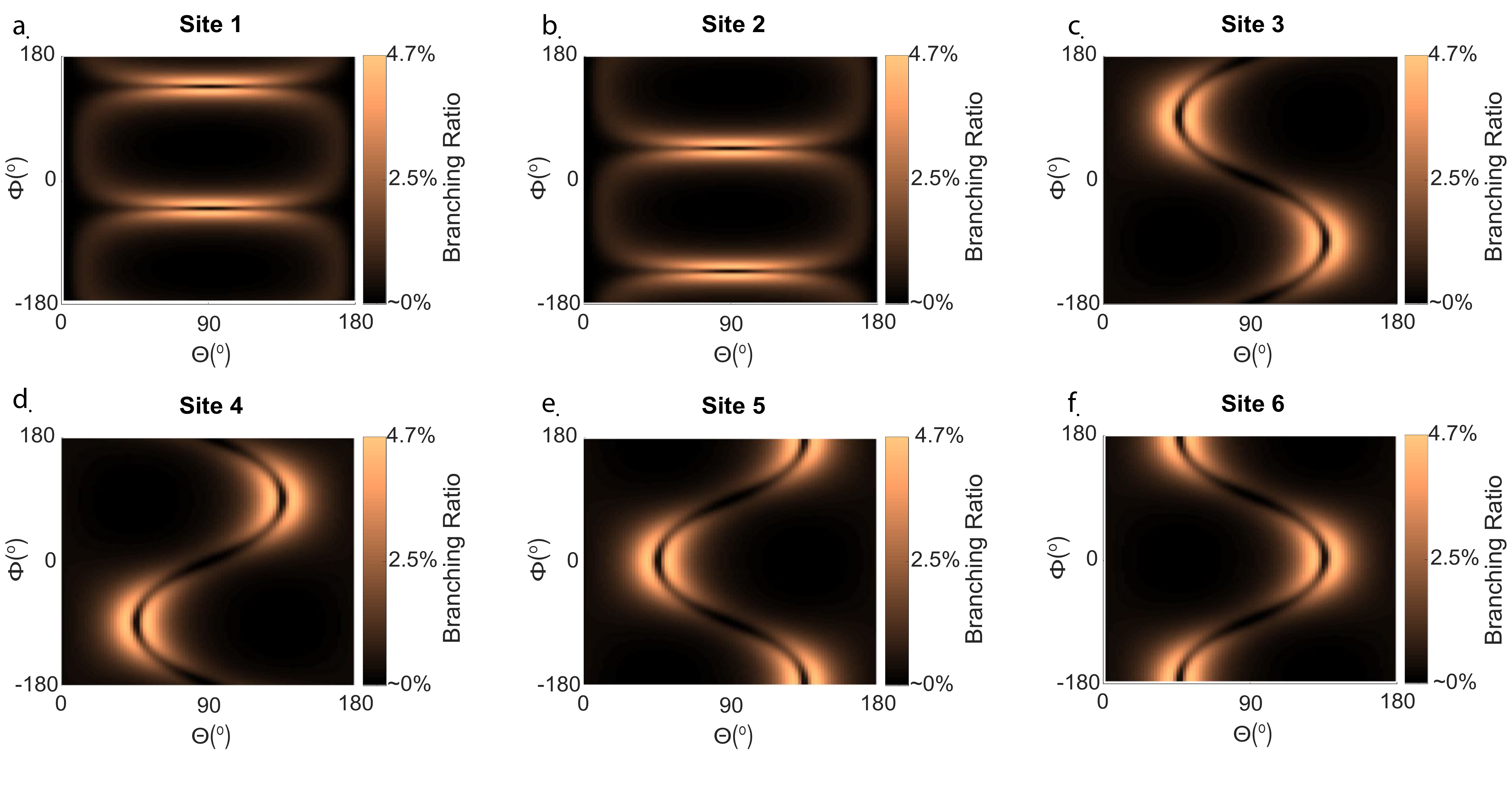}	
\caption {\textbf{a-f.}Site by site orientation map of the branching ratio R as defined in \cite{deSeze2006}. Each site reaches a maximum branching ratio of ~5\% for certain angles close to the particular local $\hat{x}$ axis}
\label{SIFig4}
\end{figure}
Due to the lower anisotropy of the hyperfine tensors, the maximum branching ratios that appear possible in Tm:YGG is ~5\%, much lower than the optimal case of ~25\% in Tm:YAG.

%\bibliographystyle{ieeetr}
%bibliography{YGGOrientBib}	